\documentclass[11pt]{article}
\usepackage{fullpage}
\usepackage{color}
\usepackage{graphicx}
\usepackage{epsf}
\usepackage{epsfig}
\usepackage{mathtools} 
\usepackage{enumitem} 
\usepackage{subfig} 
\usepackage{rotating} 
\usepackage{multirow} 
\usepackage{amsthm}
\usepackage{latexsym}
\usepackage{amssymb}
\usepackage{amsmath}
\usepackage{subfig}
\usepackage{algorithm}
\usepackage{algpseudocode}
\usepackage{wrapfig}
\usepackage{ifthen}
\usepackage{caption}
\usepackage{pgfplots}

\tikzstyle{internal} = [draw, fill, shape=circle]
\tikzstyle{external} = [shape=circle, draw]
\tikzstyle{square}   = [draw, fill, rectangle, inner sep=5pt]
\usepackage{tikz}
\usetikzlibrary{arrows,backgrounds,positioning,calc,fit,shapes.geometric,decorations.pathreplacing,decorations.markings}
\graphicspath{{img/}{./}} 

\newcommand{\newfontobj}[2]{
	\newcommand{#1}[1]{
		\expandafter\def\csname##1\endcsname{{#2 ##1}}}}

\newfontobj{\class}{\rm} 

\def\final{1}  
\def\iflong{\iffalse}
\ifnum\final=0  
\newcommand{\jnote}[1]{{\color{red}[{\tiny \textbf{Jake:} \bf #1}]\marginpar{\color{red}*}}}
\newcommand{\jycnote}[1]{{\color{purple}[{\tiny \textbf{Jin-Yi:} \bf #1}]\marginpar{\color{purple}*}}}
\newcommand{\dnote}[1]{{\color{olive}[{\tiny \textbf{Dani:} \bf #1}]\marginpar{\color{olive}*}}}
\newcommand{\knote}[1]{{\color{blue}[{\tiny \textbf{Ken:} \bf #1}]\marginpar{\color{blue}*}}}
\else 
\newcommand{\jycnote}[1]{}
\newcommand{\jnote}[1]{}
\newcommand{\dnote}[1]{}
\newcommand{\knote}[1]{}
\fi

\class{PSPACE}
\class{L}
\class{SL}
\class{BPL}
\class{RL}
\class{NC}
\class{ZPL}
\class{NPSPACE}
\class{ASPACE}
\class{NL}
\class{coNL}
\class{EXP}
\class{NEXP}
\class{coNEXP}
\class{NE}
\class{E}
\class{AM}
\class{MA}
\class{NP}
\class{DNP}
\class{UP}
\class{P}
\class{RP}
\class{BPP}
\class{ZPP}
\class{EXPSPACE}
\class{coNP}
\class{coRP}
\class{coAM}
\class{IP}
\class{PCP}
\class{AC}
\class{TC}
\class{NC}
\class{MIP}

\class{BP}

\DeclareMathOperator{\cw}{cw}
\DeclareMathOperator{\pw}{pw}
\DeclareMathOperator{\bw}{bw}

\DeclareMathOperator*{\argmin}{arg\,min}

\newcommand{\MM}{\mathcal{M}}

\newcommand{\PH}[2]{
	\ifthenelse{\equal{#1}{p}}{\Sigma_{#2}^P}{
		\ifthenelse{\equal{#1}{s}}{\Pi_{#2}^P}{
			\ifthenelse{\equal{#1}{d}}{\Delta_{#2}^P}{
				\ifthenelse{\equal{#1}{t}}{\Theta_{#2}^P}{ERROR}
			}
		}
	}
}

\newtheoremstyle{example}{\topsep}{\topsep}%
{\normalfont \small}   
{}    
{\bfseries}     
{}
{\topsep}
{}
\theoremstyle{example}

\title{A Uniformly Random Solution to Algorithmic Redistricting}
\author{Jin-Yi Cai\and Jacob Kruse\and Kenneth Mayer\and Daniel P. Szabo}

\begin{document}
	
	\maketitle
	\begin{abstract}
		\dnote{Example note}\jycnote{Example note}\jnote{Example note}\knote{Example note}
		The process of drawing electoral district boundaries is known as political redistricting. Within this context, gerrymandering is the practice of drawing these boundaries such that they unfairly favor a particular political party, often leading to unequal representation and skewed electoral outcomes.
		
		One of the few ways to detect gerrymandering is by algorithmically sampling redistricting plans. Previous methods mainly focus on sampling from some neighborhood of ``realistic' districting plans, rather than a uniform sample of the entire space.
		We present a deterministic subexponential time algorithm to uniformly sample from the space of all possible $ k $-partitions of a bounded degree planar graph, and with this construct a sample of the entire space of redistricting plans. We also give a way to restrict this sample space to plans that match certain compactness and population constraints at the cost of added complexity. The algorithm runs in $ 2^{O(\sqrt{n}\log n)} $ time, although we only give a heuristic implementation. Our method generalizes an algorithm to count self-avoiding walks on a square to count paths that split general planar graphs into $ k $ regions, and uses this to sample from the space of all $ k $-partitions of a planar graph.
	\end{abstract}
	
	\section{Introduction}
	
	Political districting is a central problem in many electoral systems, and gerrymandering can result in voter disenfranchisement as well as unfair elections. Gerrymandering is the act of creating electoral districts in a manner that 
	unfairly favors a particular political party. One of the proposed solutions to this practice is algorithmic redistricting, where possible districting plans are sampled automatically. Although there is much legal discourse about the applicability of algorithmic redistricting to existing gerrymandering cases, the problem is of interest in a purely mathematical sense, and an algorithm to generate truly uniform districting plans would have multiple important applications, such as evaluation of other sampling methods and insight into the structure of the space. 
	
	One of the most fruitful approaches to combating gerrymandering is known as \textit{outlier analysis}. This involves sampling an ensemble of districting plans, and seeing whether the current plan is an outlier in the space of all these plans. This method has been successful in numerous district and state supreme court cases, although in (\textit{Rucho v. Common Cause}, 2019) the U.S. Supreme Court decided it was too difficult for a federal court to decide how much of an outlier is enough to detect gerrymandering. Nonetheless, outliers in metrics such as 
	\jnote{what is the particular metric used as race? like percent black? need to bridge the gap between "metric" and "race", which is more of a concept rather than a specific metric for that concept}\dnote{I want some metric that measures segregation- I think percent black would do that, but don't know the best way to put it}\jnote{Ken would know best on this}
	diversity are evidence of racially segregated districts, which has been considered by the Supreme Court to be a violation of the Voting Rights Act.
	
	Many methods to produce ensembles via algorithmic redistricting for the purpose of outlier analysis have been proposed. For example there are methods that 
	restrict the sample space to particular district shapes, such as in \cite{voronoi} where they randomly assign district centers, construct Voronoi cells based on some distance metric, and make minor adjustments to balance population. This method treats hard compactness constraints as a part of the problem definition, but as 
	the authors point out, optimizing compactness greatly restricts the space of possible districting plans. Another method is outlined in \cite{arcs}, where a divide and conquer based strategy is used to recursively split a district and then make slight adjustments to balance population. It also greatly restricts valid district shapes to (approximately)
	\jnote{it restricts the valid district shapes bc they must follow arcs? Unlcear. Also unclear how this ties into the voronoi component comment}\dnote{Is this better? They recursively split the graph via arcs (curved lines on a map) following some distribution that has some compactness maximizing part via Voronoi components, but I don't know the details}\jnote{That's clearer to me}
	follow geometric arcs, and also has a compactness maximizing Voronoi component.
	
	Some more classical approaches to generating district plans include formalizing the problem as a linear program \cite{1970LP}. Their method has two phases, 1) they first generate 
	feasible (with respect to population constraints) districts and then 2) formalize an optimization problem to combine these districts into a valid, compact districting plan. Mehrota et al. \cite{ImprovedLP} improve on this method, however, they still only optimize some heuristic and only aim to create a single ``fair" districting plan.
	
	Other algorithms include Yamada's heuristic minimax spanning forest algorithm \cite{spanningLP}, where they choose a set of $ k $ roots and then find a spanning forest on these that optimizes some integer program representing some feasibility constraints. Even approaches from the genetic literature \cite{genetics1, genetics2} are suggested, most of which use evolutionary updates similar to the MCMC methods below, which can take factors such as population, contiguity, and compactness into account. Methods from statistical physics have been applied as well. For example, Chou and Li \cite{pottsredist} model the adjacency graph as a $ k $-state Potts model, of which the steady state 
	models a possible districting plan.
	
	Beyond these are a slew of Markov Chain Monte Carlo (MCMC) methods. MCMC methods construct a random walk on some space of districting plans,
	and sometimes offer mixing theorems to guarantee convergence to some distribution on this space. Most begin with some districting plan and then perform some ``flips" on the boundaries such that the population deviation between districts is kept low and district-level compactness is kept high. See \cite{ReCom, mcmc1, mcmc2} for examples of applications of MCMC in redistricting. Although these methods can efficiently satisfy some of the necessary constraints for districting plans, they are unable to converge fast enough to their stationary distribution \cite{Frieze_MC}. Due to this they can only sample some local deviations in the space of all districting plans, which is wrought with local maxima (with regards to the constraints) that MCMC methods cannot explore in a feasible amount of time. 

	These MCMC methods do not always converge to the uniform distribution, and those that do don't have reasonable mixing times.
	A very recent result of \cite{FriezePegden} however is able to sample from the uniform distribution in subexponential time by proving bounds on the mixing time of the Glauber dynamics of grid-like graph.
	Their results show that the mixing time is fixed parameter tractable in the \emph{bandwidth} of a given graph $ G $, which we will define in section \ref{sec:param}. One important difference is their assumption that the size of the partition, $ k, $ is linear in $ n $. This makes it impractical for most redistricting applications. \dnote{Add something like that says it's a good result bc it is pretty cool} 

	
	One of the many challenges in the redistricting process is the enormous number of possible redistricting plans. For example, we calculated the number of possible bipartitions of a map with 421 nodes, without population and compactness constraints, to be $ \approx5.53\times 10^{80} $, and $ 2.81\times 10^{35} $ with discrete perimeter and loose population constraints. 
	The scale of these values means that for a sample to be representative of the space of valid districting plans, it not only needs to be large enough to make powerful statistical claims, but also unbiased. Restricting 
    the sample to some local space gives inherent bias to the sample, even if the districting plan in question comes from this neighborhood.
	
	Our solution to this challenge is to approach the problem with a provably uniform sample from the space of all possible districting plans. When defining this space, we can efficiently enforce compactness constraints, but not population. We then use rejection sampling to loosely satisfy the other constraints needed of a valid districting plan.\jnote{Your commented out line here seems like a useful thing to include.} \dnote{I added it, but I'm not sure if we can still say it has remained elusive}\jnote{because of the other recent paper you describe above? If so, that
's really only for two districts, so maybe you could say something like "for realistic redistricting scenarios"}\dnote{I don't know, do we really get realistic districting scenarios? It's definitely more realistic than the other one-- I think elusive in the field is fine, it doesn't necessary mean nobody has gotten close} This allows us to grasp the entire space of valid districting plans, a distribution that has so far remained elusive in the field.

	The significant challenges to the applicability of this approach are pointed out in \cite{ReCom}. First, the problem as defined here is 
	$ \NP $-hard, meaning we cannot find a polynomial sampler unless $ \NP=\RP $ \cite{hardness}. For this reason, we only present a subexponential time algorithm. Second, any uniform sample on the entire space is exponentially wrought with constraint violating partitions. These are partitions that are not compact, such as space filling curves, or partitions that do not satisfy population constraints. 
	This issue we address by carefully defining the space we sample from. We give not only an algorithm to uniformly sample connected partitions of a planar graph, but one to to uniformly sample from the space of \emph{valid} partitions, where certain compactness and population constraints must be satisfied. The sample can then be further restricted via rejection sampling or local corrections to reach the strict constraints present in redistricting applications.
	\dnote{Maybe mention that we don't consider the splitting penalties for counties?}\jnote{Could be good to note in the discussion section, along with any other common criteria that aren't included in the method.}
	
	
	One unique benefit of this work is the generation of exact counts of possible partitions. It has been known that the number of valid districting plans is enormous, but the sheer scale of this enormity did not have any computational evidence. Our experiments put this enormity into perspective, as well as the ratio of plans that are valid with respect to population and compactness constraints. For example, we give experimental evidence that the number of contiguous partitions with a given discrete perimeter grows exponentially with the perimeter, showing that the space of all connected partitions truly is wrought with so-called space filling curves.
	
	Our results are similar to those of Frieze and Pegden \cite{FriezePegden}, as both of the proposed methods generate a uniform sample of districting plans in subexponential time. However, there are a few key differences. First, their results hold in the regime where the size of the partition $ k $ is linear in $ n $, while ours hold even when $ k $ is constant. The algorithmic redistricting application falls under the regime where $k$ is constant. 
	The other difference is in the graph parameter appearing in the running time. The algorithm given in \cite{FriezePegden} runs in time exponential in the bandwidth of the dual graph, while ours is exponential in the cutwidth of the primal. Here the primal graph is the graph where the faces are precincts, and the dual where the precincts are nodes. The relation between these parameters is explored in Section \ref{sec:param}. As we will show, there are better bounds for the cutwidth of a bounded degree planar graph, meaning our algorithm runs in $ 2^{O(\sqrt{n}\log n)} $ on any such graph.
	
	Potential applications of the method presented here include evaluation of other, existing sampling methods, as well as the usual statistical method of evaluating an existing districting plan with outlier analysis. 
	Many sampling methods, such as \cite{ReCom}, sample have been sampling from wider and wider distributions,
 but there has so far been almost no way to evaluate how close the sample is to a uniform sample. 
	We offer an implemented benchmark distribution that is truly uniform, even when certain constraints are only handled via rejection sampling.
	This method could also be combined with other existing methods. MCMC methods, for example, start from some initial map and efficiently explore some local neighborhood of this map. Choosing the initial map from the uniform distribution would give the key global component to these sampling methods.
	
	\section{Background}
	
	\subsection{Combinatorial Concepts}
	
	Some of the methods described in this paper are based on common topics in combinatorial optimization and enumerative combinatorics. As we want this method to be as widely understandable as possible, we begin by introducing some of these definitions.
	
	The first and most important definition is that of a graph $ G=(V,E) $, defined by a set of vertices $ V=\{v_1,v_2,\ldots, v_n\} $ and a set of edges $ E $ between these vertices. If the graph is directed, these edges are ordered pairs of vertices $ (v_i, v_j) $, and if it is undirected then they are unordered pairs $ \{v_i, v_j\} $. We will only deal with undirected graphs, and therefore abuse the notation $ (v_i, v_j) $ for edges of undirected graphs as well. We will also use the notation $A\triangle B$ for the symmetric difference between sets $A$ and $B$.
	
	A graph $ G $ is \emph{planar} if it can be drawn into a plane without any crossing edges. Such a drawing is called a combinatorial embedding, where the nodes $ V $ are given positions and a set of \emph{faces} $ F $ arise, where faces are cycles that contain no other edge in the embedding. It's easy to see that this definition of planar is then equivalent to the existence of another type of combinatorial embedding, a spherical embedding, where the graph is drawn onto the surface of a sphere without any crossing edges. When viewing geographical entities by their adjacency graphs, we are in fact considering a spherical embedding onto the surface of the earth.
	
	For a planar graph $G=(V,E)$, the \emph{dual} of $G$ is $G^*=(F,E')$, where $(f,f')$ is an edge whenever faces $f$ and $f'$ share an edge.
	\dnote{Probably don't need this entire paragraph from here. Or add the minor free characterization, because it appears in a lot of the cutwidth and bandwidth papers we cite}
	There are countless beautiful results about planar graphs in graph theory. One of the oldest results is that of Euler, who proved any planar graph satisfies $ |V|+|F|-|E| = 2 $. 
	Another important concept about planar graphs is what data structure to use to best represent them. The na\" ive solution of storing the set of nodes and edges along with a mapping of nodes to positions does not allow for an easy way to construct the set of faces, nor does it allow a way to traverse them. The half-edge data structure solves this by storing $ G $ as a directed graph, where each undirected $ e\in E $ is replaced by two directed ``half" edges. These edges also have pointers to the next half edge in any given face, and as each edge is shared by exactly two faces, 
 we can store all of $ F $.
	
	Another background concept comes from the field of combinatorial optimization. Dijkstra's algorithm aims to solve the shortest path problem: given a graph $ G $, compute the shortest paths between some $ u $ and all $ v \in V $. It proceeds from any starting vertex and propagates via breadth-first search (for unweighted graphs), updating shortest paths for each new node using the shortest path of the previous node.\dnote{should we define BFS?} 
	
	The final introductory discussion is about a topic from enumerative combinatorics, Motzkin paths. A Motzkin path is a path traversing a 
	lattice from $ (0,0) $ to $ (0,n) $. This path is allowed to move Northeast, East, and Southeast but may never go below the $ x $ axis. The number of such paths of length $ n $ is denoted $ M_n $, the $n$th Motzkin number. It can be derived exactly from the generating function $ \MM(x) = \sum_n M_n x^n $. Using the self similar properties of Motzkin paths we see for any $ n>1 $, $ M_n $ satisfies the recurrence $ M_n=M_{n-1} + \sum_{n=0}^{n-2} M_i M_{n-2-i} $. Plugging the recurrence
	into the generating function and collecting like terms, we see $ \MM(x) = 1 + x\MM(x) + x^2 \MM(x)^2 $. This can be solved to derive
	\[ \MM(x) = \frac{1-x-\sqrt{(1+x)(1-3x)}}{2x^2}. \]
	The differential properties of this form of $ \MM(x) $ show the $ M_n $ also satisfy the recurrence
	\[ M_n = \frac{2n+1}{n+2}M_{n-1} + \frac{3n-3}{n+2} M_{n-2}. \]
	Asymptotic analysis of this recurrence shows
	\[ M_n \sim \dfrac{1}{2\sqrt{\pi}} \left( \frac{3}{n} \right)^{3/2} 3^n \quad \text{as }n\to \infty, \]
	of which we can take away $ M_n = O(3^n) $.
	
	\subsection{Bandwidth, Cutwidth, and Graph Parameters}\label{sec:param}
	
	We now proceed by discussing some well studied graph parameters that affect the complexity of the algorithms that uniformly sample connected graph partitions. These are the bandwidth (bw), cutwidth (cw), and pathwidth (pw) of a graph $ G=(V,E) $, with $ n=|V| $. All are based on some minimum over all permutations $ \pi $ of $ V $, and satisfy some inequalities.
	
	The bandwidth is defined to be the minimum 
	over all permutations $ \pi:V\to [n] $ of the maximum difference of any two neighboring vertices, that is
	\[ \text{bw}(G) = \min_\pi \max_{(u,v)\in E} |\pi(u) - \pi(v)|. \]
	The cutwidth  is the minimum over all $ \pi $ of the maximum linear cut between vertices before some threshold and after, namely
	\[ \text{cw}(G) = \min_\pi \max_{i\in [n]} |\{(u,v)\in E : \pi(u)\leq i < \pi(v)\}|. \]
	Finally, the pathwidth has a characterization known as the vertex separation number, which is defined as the vertex cut version of cutwidth, that is
	\[ \text{pw}(G) = \min_\pi \max_{i\in [n]} |\{u\in V : \exists v\in N(u)\ \pi(u)\leq i < \pi(v)\}|. \]
	
	One way to understand these parameters is by imagining $ \pi $ as a linear layout 
	of $ G $, or an embedding of $G$ onto a line. Then the bandwidth of a layout is the maximum length of an edge, the cutwidth is the maximum number of edges crossing some line perpendicular to the layout, and the pathwidth is the maximum over each of these perpendicular lines of the minimum vertex cut needed to remove all edges crossing the line.
	
	The following well known inequalities are immediate from this perspective. For any graph $ G, $
	\begin{itemize}
		\item $ \pw(G)\leq \cw(G) \leq \Delta(G)\pw(G) $, where $ \Delta(G) $ is the maximum degree of $ G $.
		\item $ \pw(G)\leq \bw(G) $.
		\item $ \cw(G) \leq \Delta(G)\bw(G) $.
	\end{itemize}
	
	For grids, all three parameters are approximately equal at $ \sqrt{n} $. For arbitrary planar graphs, it is known $ \pw(G) = O(\sqrt{n}) $ by some induction on the Lipton-Tarjan planar separator theorem \cite{planarpathwidth}. We can then extend this to say the cutwidth of a bounded degree planar graph is $ O(\sqrt{n}) $ as well. For bandwidth, the best known upper bound on planar graphs of bounded degree is $ bw(G)\leq \frac{20n}{\log_{\Delta(G)}(n)} $ \cite{bottcherbandwidth}.
	
	\dnote{We are technically using the cutwidth of the primal graph, and they use the bandwidth of the dual. There's some stuff about these params under duality, but it's weak (although relates to some open problems about pathwidth under duality), and our application is close to a grid, which is self-dual anyways. Should I include this? We don't give any direct comparison to the \cite{FriezePegden} bandwidth parameter anyways, but duality would make this comparison even weirder.}
	
	The difficulty is that these parameters are $ \NP $-hard to compute, even for planar graphs \cite{graphparam}. Nonetheless there is a rich literature of fixed parameter tractable (FPT) algorithms and approximation algorithms for these problems. For cutwidth, even the simplest dynamic programming algorithm using Held-Karp \cite{cutwidthFPT} gives an algorithms that outputs, in $ O(n^{k-1}) $ time, a labeling with cutwidth $ \leq k $, or says no such labeling exists. The FPT algorithm for this same problem of \cite{ThilikosBodlaenderSerna} runs in $ n\cdot 2^{O(k^2)} $, and the best approximation algorithm for planar graphs gives a $ O(\log n) $ approximation \cite{cwapx}. For practical computation, branch-and-bound algorithms \cite{cwbranch} have shown potential, although having worse theoretical guarantees.\dnote{Ideally, we should probably be using branch and bound for the experiments}
	
	\subsection{Transfer Matrix Enumeration Algorithms}
	
\begin{figure}
		\centering
		\begin{tikzpicture}[->,>=stealth', circ/.style={shape=ellipse, draw=black, color=black}]
			\node[circ] (A) {$ w(10, 10) $};
			\node[circ, below=.4cm of A] (B) {$ w(9, 10) $};
			\node[circ, below left=.7cm and 2.5cm of B] (C) {$ w(8, 10) $};
			\node[circ, below left=.7cm and 1cm of C] (E) {$ w(7, 10) $};
			\node[circ, below left=.7cm and 0cm of E] (E1) {$ w(6, 10) $};
			\node[circ, below right=.7cm and 0cm of E] (E2) {$ w(7, 9) $};
			\node[circ, below right=.7cm and 1cm of C] (F) {$ w(8, 9) $};
			\node[circ, below left=.7cm and 0cm of 	F] (F1) {$ w(7, 9) $};
			\node[circ, below right=.7cm and 0cm of 	F] (F2) {$ w(8, 8) $};
			\node[circ, below right=.7cm and 2.5cm of B] (D) {$ w(9, 9) $};
			\node[circ, below=0.4cm of D] 	(G) {$ w(8, 9) $};
			\node[circ, below left=.7cm and 0cm of 	G] (G1) {$ w(7, 9) $};
			\node[circ, below right=.7cm and 0cm of 	G] (G2) {$ w(8, 8) $};
			\node[below=0cm of F2] (ldots) {$ \vdots $};
			\path[draw] (A)->(B);
			\path[draw] (B)->(C);
			\path[draw] (C)->(E);
			\path[draw] (E)->(E1);
			\path[draw] (E)->(E2);
			\path[draw] (C)->(F);
			\path[draw] (F)->(F1);
			\path[draw] (F)->(F2);
			\path[draw] (B)->(D);
			\path[draw] (D)->(G);
			\path[draw] (G)->(G1);
			\path[draw] (G)->(G2);
		\end{tikzpicture}
		\caption{The na\"ive call graph for well formed bracket expressions.}\label{fig:naive}
	\end{figure}
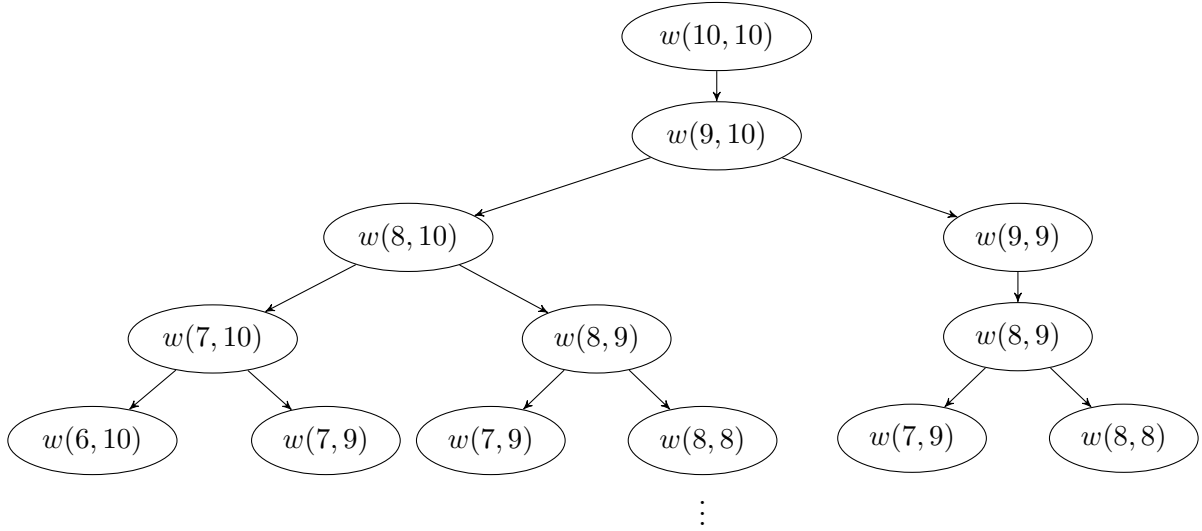

	In this section, we introduce the concept of transfer matrix algorithms for enumeration. Say we wanted to enumerate the set of all well formed bracket expressions with $ n $ opening brackets and $ n $ closing brackets $ w(n,n) $ as in \cite{transfmatr}. The function $ w $ satisfies the following recurrence for $ o $ opening brackets and $ c $ closing:
	\[ w(o,c) = \left\{
	\begin{array}{ll}
		1 & \mbox{if } o=0 \mbox{ and } c=0 \\
		0 & \mbox{if } o>c \\ 
		w(o-1,c) & \mbox{if } o=c>0 \\
		w(o-1,c)+w(o,c-1) & \mbox{otherwise}
	\end{array}
	\right. \]
	This recurrence gives a na\" ive recursive implementation of how to enumerate $ w(n,n) $, however this implementation would run in exponential time ($ O(2^n) $). The first few layers of the call graph is shown in Figure \ref{fig:naive}.
	
	The natural improvement is to use a dynamic programming (DP) algorithm. These algorithms utilize \emph{memoization}, where repeated calls to the same function with the same input are stored in memory to speed up calculation. Thus, when calculating $ w(n,n) $ we need only calculate each $ w(o,c) $ once, to enumerate all the expressions in $ O(n^2) $ time. 
	
	\begin{figure}
		\centering
		\begin{tikzpicture}[->,>=stealth', circ/.style={shape=ellipse, draw=black, color=black}]
			\node[circ] (A1) {$ w(10, 10) $};
			\node[circ, below=.4cm of A1] (B1) {$ w(9, 10) $};
			\node[circ, below left=.7cm and 1cm of B1] (C1) {$ w(8, 10) $};
			\node[circ, below right=.7cm and 1cm of B1] (C2) {$ w(6, 10) $};
			\node[circ, below=.4cm of C1] (D1) {$ w(7, 10) $};
			\node[circ, below=.4cm of C2] (D2) {$ w(7, 10) $};
			\node[circ, below right=.7cm and 1cm of D1] (E2) {$ w(8, 9) $};
			\node[circ, below left=.7cm and 0cm of D1] (E1) {$ w(6, 10) $};
			\node[circ, below right=.7cm and 0cm of D2] (E3) {$ w(8, 9) $};
			\node[below=0cm of E2] (ldots) {$ \vdots $};
			\path[draw] (A1)->(B1);
			\path[draw] (B1)->(C1);
			\path[draw] (B1)->(C2);
			\path[draw] (C1)->(D1);
			\path[draw] (C1)->(D2);
			\path[draw] (C2)->(D2);
			\path[draw] (D1)->(E1);
			\path[draw] (D1)->(E2);
			\path[draw] (D2)->(E1);
			\path[draw] (D2)->(E2);
			\path[draw] (D2)->(E3);
		\end{tikzpicture}
		\caption{The DP and TM call graph for well formed bracket expressions.}\label{fig:DP}
	\end{figure}
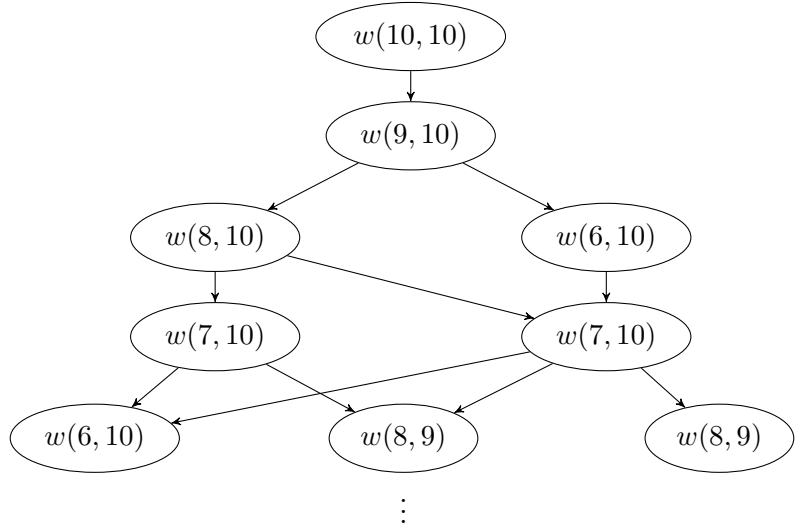
	
	Transfer matrix (TM) algorithms \cite{transfmatr} are ideal for enumerating combinatorial objects, even if they have exponentially many configurations. It is a technique similar to dynamic programming (DP), in that it utilizes some form of memoization. A TM algorithm separates a given problem into a series of layers $ j=1,2,\ldots, m $, and states $ S_j $ for each layer. These subproblems are such that each state $ s_{i,j} \in S_j $ can be expressed as a linear combination $ f_j $ of states in the previous layer $ S_{j-1} $. Using this we need only express the multiplicities of each state, as $ S_m = f_m(S_{m-1}) = f_m(f_{m-1}(S_{m-2})) = \ldots = f_m(f_{m-1}(\ldots f_2(S_1)\ldots )) $. Assuming $ |S_m|=|S_1|=1 $, because each $ f_j $ is linear we see $ f=f_2\circ f_3 \circ \ldots \circ f_m $ is linear as well, and therefore just some multiple of the starting state plus a constant. Here, similar to DP, each state $ s_{i,j} $ is getting contracted into a multiple, rather than na\"ively performing the calculation multiple times. The call graph for both DP and TM implementations are shown in Figure \ref{fig:DP}.
	
	A key difference however is that TM algorithms have a strict order on the layers, and this allows for lower space complexity as well as other tricks presented in \cite{transfmatr}. A DP call graph can be any rooted acyclic graph, while a TM algorithm ensures that the resulting call graph is graded. Despite their advantages for enumeration, TM algorithms have some disadvantages for other computation. A common motif in computer science is that counting \emph{is} sampling, however this does not hold exactly for TM algorithms. For DP algorithms we can store a dictionary entry with the solution to each subproblem, and this allows sampling simply by stepping through the full call graph with probabilities proportional to the counts. For TM algorithms we are restricted to storing only one layer at a time, which means sampling can only be done dynamically. For \emph{clean} TM algorithms, where there are no states in $ S_{j} $ that are not referenced by any state in $ S_{j+1} $, dynamic sampling is equivalent to a uniform sampling of paths from $ S_j $ to $ S_1 $. Dynamic sampling of an unclean system however can result in a prohibitively large percentage of samples being lost.
	It is difficult to clean an algorithm, although there exist methods such as signature trimming to do so.
	
	In this work, we present an unclean TM algorithm that is implemented as a DP algorithm to allow for sampling. Although we did implement dynamic sampling, the allocation of additional permanent memory to hold the full call graph proved more efficient than cleaning the algorithm.
	
	\section{Sampling Self-Avoiding Walks}\label{sec:alg}
	Our algorithm samples self-avoiding walks (SAWs) on the edges of a planar graph that split the graph into $ k $ connected components. It is based off of a matrix enumeration algorithm to count self-avoiding walks on an $ L\times M $ grid discovered in \cite{grid_walks}, which we repeat here for clarity. 
	
	\subsection{Bipartitioning a Grid}\label{grid}
\begin{figure}[!tbh]
		\centering
		\includegraphics[scale=1]{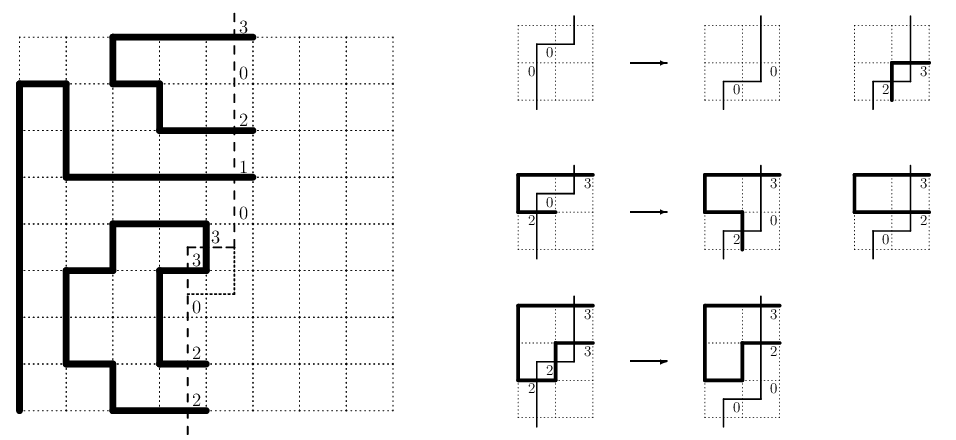}
		\caption{A partial self avoiding walk on a grid, with possible update steps on the right. Figure from \cite{grid_walks}.}\label{traversal}
	\end{figure}

	The simplest version of the problem is when 
	we consider a single path crossing a grid. We can then decompose the problem into $ LM $ layers, and $ O(3^L) $ possible states at each layer. We assume without loss of generality that $ L<M $. The layers are defined by steps along the faces of the grid, starting with bottom left, moving up along the column, and then repeating one column to the right, and so on. Given a layer, we call the boundary (the dotted line in Figure \ref{traversal}) a \textit{frontier} $ \delta T_j $ for $ 0\leq j < LM $, where $ \delta T_j $ is an ordered set of edges on the grid. The states are then given by labels $ \sigma_i $ of edges $ i\in \delta T_j $ along a frontier by symbols $ 0,1,2, $ or $ 3. $
	
	A given frontier along with a labeling then gives us a subproblem for our transfer matrix algorithm. This problem is to count the number of partially completed walks intersecting the frontier in accordance with the labeling $ \sigma_i $. The symbols along a frontier have the following relationship with partial SAWs:
	\begin{itemize}
		\item If $ \sigma_i = 0 $, no partial SAW edge crosses grid edge $ i $.
		\item If $ \sigma_i = 1 $, the partial SAW crosses the frontier at edge $ i $, and this path is directly connected outside boundary.
		\item If $ \sigma_i = 2 $, the partial SAW crosses the frontier at edge $ i $, and then loops back at a paired edge $ i' $ such that $ \sigma_{i'}=3 $.
		\item If $ \sigma_i = 3 $, the partial SAW crosses the frontier at edge $ i $, and then loops back at a paired edge $ i' $ such that $ \sigma_{i'}=2 $.
	\end{itemize}
	
	The pairing of $ 2 $'s and $ 3 $'s can be realized by viewing them as balanced parentheses, or by considering a Motzkin path where a $ 2 $ is a northeast step, a $ 3 $ a southeast step, and a $ 0 $ or $ 1 $ a horizontal step. There is then a clear matching between northeast and southeast steps. With this matching in mind, the symbols $ 3 $ and $ 2 $ can be better thought of as pointers to the matched edges. Define the $ \sigma $-dependent function $ p $ such that $ \sigma_i $ is paired with $ \sigma_{p(i)} $, where $ p(i)=i $ when $ \sigma_i=0,1 $.
	
	Notice then that each labeling $ \sigma $ of $ \delta T_j $ corresponds to a Motzkin path, so the number of states at a given layer is no more than the number of Motzkin paths, of which there are $ O(3^{|\delta T_j|}) = O(3^L) $. 
	
	For each labeling $ \sigma $ of $ \delta T_j $, it is simple to enumerate all the possible labelings $ \sigma' $ of $ \delta T_{j+1} $ that any partial SAW with boundary $ \sigma $ can evolve into. The symmetric difference between $ \delta T_j $ and $ \delta T_{j+1} $ is at most 4 edges, 2 of which are labeled in $ \sigma $ and 2 that are labeled in $ \sigma' $. 
	
	In Figure \ref{traversal}, we can see some of the possible updates. They are summarized in Table \ref{update-table}. The notation $ \widehat{00} $ refers to a main path meeting a $ 2 $ or a $ 3 $, in which case the matched $ \sigma_{p(i)}=2\text{ or }3 $ is updated to $ \sigma_{p(i)}' = 1 $. The notation $\overline{00}$ represents when two edges labeled with $ \sigma_i=\sigma_{i+1} $ meet. In this case if $ \sigma_i=2 $ then $ \sigma_{p(i)}' = 2 $, and the new matching $ p' $ has $ p'(p(i)) = p(i+1) $. Otherwise if $ \sigma_i=3 $, $ \sigma_{p(i+1)}' = 3 $, and the new matching $ p' $ has $ p'(p(i+1)) = p(i) $. Finally the notation Res refers to a final result, where we terminate a free end and output our result, assuming no other edges are occupied by $ 2 $s or $ 3 $s.
	
	\begin{table}[!tbh]
		\begin{center}
			\renewcommand{\arraystretch}{1.2}
			\begin{tabular}{|c|ccc|ccc|cc|cc|}  \hline  \hline
				\raisebox{-1.5mm}{Bottom}\raisebox{-0.5mm}{\Large $\backslash$}\raisebox{0.5mm}{Top}
				&\multicolumn{3}{c|}{0} &\multicolumn{3}{c|}{1} & \multicolumn{2}{c|}{2}
				& \multicolumn{2}{c|}{3}  \\
				\hline
				0  &  $00$   & $23$  &
				&  $01$   & $10$ & Res
				&  $02$   & $20$
				&  $03$   & $30$    \\    \hline
				1  & $01$   & $10$ & Res
				&  & &
				&  \multicolumn{2}{c|}{$\widehat{00}$}
				&  \multicolumn{2}{c|}{$\widehat{00}$}  \\  \hline
				2  &  $02$   & $20$ &
				&  & $\widehat{00}$ &
				&  \multicolumn{2}{c|}{$\overline{00}$} & &
				\\  \hline
				3   & $03$   & $30$ &
				&  & $\widehat{00}$ &
				&  \multicolumn{2}{c|}{$00$}
				&  \multicolumn{2}{c|}{$\overline{00}$}   \\
				\hline \hline
			\end{tabular}
		\end{center}
		\caption{
			The various `input' states and the `output' states which arise as the
			boundary line is moved in order to include one more vertex.
			Each panel contains up to three possible `output' states or other allowed
			actions. The table is from \cite{grid_walks}.}\label{update-table}
	\end{table}
	
	With the given transitions, we maintain counts $ c_{\sigma, j} $ for each labeling $ \sigma $ of $ \delta T_j $. The total number of SAWs is then $c_{f, LM}$, where $f$ is the empty function.
	
	\subsection{Partitioning the Dual Graph}
	We generalize this method to arbitrary planar graphs, and to general $ k $-partitions. As input, we are given the primal graph $ G_0=(V_0, E_0) $ with a spherical embedding and faces $ F_0 $, with some outer face $\tau$. In our gerrymandering applications, this is the graph where the faces are census geographies, e.g., census tracts, and the edges are the boundaries between these building blocks. From this we define the dual graph $ G=(V,E) $ with $ n=|V|, m=|E| $, along with a spherical embedding of $ G $ with a particular outer face $ \tau $. Let $ F $ be the set of faces in the planar embedding of $ G $, and $ \ell=|F|-1 $ to account for the outer face. Assume additionally that the size of any non-outer face is bounded by some constant $ \beta $.
	
\begin{figure}[!tbh]
		\centering
		\resizebox{0.3\textwidth}{!}{%
			\begin{tikzpicture}
				\draw (0,0) circle (1);
				\draw[color=blue] (.5, .866) -- (.5, 0);
				\draw[color=blue] (.5, 0) -- (.866, -.5);
				\draw[color=blue] (.5, 0) -- (0, -1);
				\draw[color=blue] (-.5, -.866) -- (-.5, 0);
				\draw[color=blue] (-.5, 0) -- (-.866, .5);
				\draw[color=blue] (-.5, 0) -- (0, 1);
				\draw[color=red, dashed] (-.289, .957) .. controls (.25, 0) .. (0.289, -.957);
			\end{tikzpicture}
		}%
		\caption{An example of splitting a map (represented by a disc) into 5 parts via 2 SAWs (in blue) with 2 splits.
			Also shown is a potential frontier which would have two edges labeled $ 1 $ as a red dashed line. These are the edges where the frontier intersects the SAW.
		} \label{fig:split_ex}
	\end{figure}
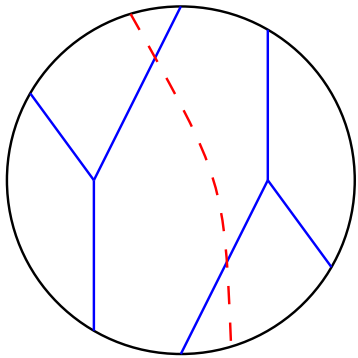

	We then seek to generate a partitioning of the vertices $ V $ of $ G $ into $ k $ mutually distinct subsets that induce connected components in $ G $. This can be done by sampling a set of mutually self-avoiding paths with no loops and a fixed number of splits/merges in $ G_0 $. The number of splits/merges, where three edges of a path share a single vertex, is inversely proportional to the number of paths. For example, to partition a graph into 5 districts, one possible solution would have two paths with two splits/merges (Figure \ref{fig:split_ex}). 
	
	In general, if we have $ i $ splits for $ 0\leq i \leq k-1 $, partitioning into $ k $ districts would yield $ k-i-1 $ paths. This adds an additional dimension to our algorithm: we must now store counts $ c_{\sigma, j, h} $ for $ 0\leq h \leq 2(k-1) $. The value $ 2(k-1) $ comes from the fact that each path must ``enter" as well as ``exit," and that the $ k $th partition is already fixed by the first $ k-1 $. We say a path can ``enter" from a split or from some edge in $ \tau $, and ``exit" into a merge or some edge in $ \tau $. As one can see in Figure \ref{fig:split_ex}, there are a total of $ 9 $ possible states.
	
	Let $ \{T_j\}_{j=1}^\ell $ be a sequence of increasing subsets of $ F $ such that $ T_1\subset T_2\subset\ldots\subset T_\ell=F $, where the symmetric difference between $ T_j $ and $ T_{j-1} $ is a single face. This sequence is just an ordering of the faces $ F $. As we go along this sequence of subsets, we can algorithmically define a growing sequence of frontiers $ \delta T_j $. With each additional face $f$, let $ f_j = T_{j-1} \triangle (f\setminus \tau) $, and let $ \delta T_j=\delta T_{j-1}\triangle f_j $. In other words, add all edges of $ f $ that aren't in $ \tau $ or $\delta T_{j-1}$ to $ \delta T_{j-1} $ to create $ \delta T_j $, and then remove any of the edges of $ f $ that were in $ \delta T_{j-1} $. We additionally assume, for algorithmic simplicity, that this frontier is connected for every $ j $, or that the edges of $ \delta T_j $ form a single path. This requirement is not completely necessary, and not every planar graph admits such a traversal. \dnote{Should expand on this- the bad case in Frieze Pegden happens to not have a short contiguous traversal}
	
	This sequence of the faces of $ G $ corresponds to an ordering of the vertices of $ G_0 $, the primal graph. The length of a frontier $\delta T_j$ is the number of edges between $T_j$ and $F\setminus T_j$, which is the same as the number of edges between these node sets in the dual graph. The minimum over all orderings of the maximum length of a frontier $ \kappa $ is then exactly the cutwidth of $G$, $\cw(G)$.
	
	Assume for now we are given such a sequence of frontiers, Section \ref{sec4} is dedicated to the process of constructing this sequence \jnote{I'm confused by this transition...assuming we want such a sequence, or that we already have it?}\dnote{I think it's clear enough, right?}\class{The transition between the clauses of this sentence is unclear to me.}. Then we can assign labelings $ \sigma $ to the edges of $ \delta T_j $, where $ \sigma $ maps the edges of $ \delta T_j $ to $ \{0,1,2,3\} $ such that $ \sigma $ forms a Motzkin path, as in section \ref{grid}. Then we again have the property that each $ \delta T_j $ along with a labeling $ \sigma $ can correspond to a count in our matrix enumeration algorithm.
	
	In this case where not all faces are quares, the update rules become slightly harder to describe. 
	Given a face $ f_j $ to be added to $ T_j $ to create $ T_{j+1} $, there could be $ 0,1 $ or $ 2 $ edges labeled in $ \delta T_j $, and at most $ \beta-1 $ edges to be labeled in $ \delta T_{j+1} $.
	
	Let $ f_j^0 = f_j \cap \delta T_j $ and $ f_j^1 = f_j \cap \delta T_{j+1} $. Then the update rules are as follows for an incoming labeling $ \sigma $ of $ \delta T_j $ in state $ h $:
	\begin{itemize}
		\item If there are $ 0 $ $ \sigma $-labeled edges in $ f_j^0 $, then there can be either no labeled edges in $ f_j^1 $, or a new $ 2 $-$ 3 $ pair between any of the $ \binom{|f_j^1|}{2} $ ordered pairs of edges in $ f_j^1 $. Note that if $ |f_j^1|<2 $, this cannot be done. We also allow, if any edge of $ \tau $ is in $ f $ and $ h<2(k-1) $, an ``entry" where any single $ i\in f_j^1 $ can have $ \sigma_{i}' = 1 $ and increase $ h $.
		
		\item If there is a single labeled edge in $ f_j^0 $, then it can either continue to any of the edges in $ f_j^1 $ with the same label, or if the label happens to be $ 1 $ we allow for a split as well. A split chooses any of the $ \binom{|f_j^1|}{2} $ pairs of edges in $ f_j^1 $ and labels them both as $ 1 $, increasing the state $ h $. Thus we only allow a split if $ h<2(k-1) $.
		
		\item If there are 2 labeled edges $ \sigma_i $ and $ \sigma_{i'} $ in $ f_j^0 $, then the update rules are for the most part the same as in Table \ref{update-table}. The only exception is when $ \sigma_i=\sigma_{i'}=1 $ in which case there are two possibilities. If $ h<2(k-1) $, we allow a merge, where any single $ i\in f_j^1 $ can have $ \sigma_{i}' = 1 $ and the state is incremented. We also, for any $ h $, allow the case where $ \sigma_{i}'=0 $ for all $ i\in f_j^1 $, without increasing the value of $ h $. This handles cases where two ``entries" meet-- for this reason we keep the term entry and exit in quotations, as it is not only traversal dependent, but also inconsistent.
		
		A consequence of this approach is that it allows ``lollipop loops" where two $ 1 $s that came from a split later merge. This results in partitions that are enclosed entirely by another partition, something that does not result in a legal redistricting plan. For the purpose of sampling, such cases are handled via rejection sampling.
	\end{itemize}
	
	Given these update rules we can pair the counts $ c_{\sigma, j, h} $ with some $ c_{\sigma', j+1, h'} $ for our transfer matrix algorithm.
	
	\subsubsection{Complexity}
	
	The running time of this algorithm is essentially the number of possible triples of $ (\sigma, j, h) $. We know $ j $ can take $ \ell $ possible values, $ h $ can take $ 2(k-1)+1 $, however the number of possible assignments $ \sigma $ depends on $ |\delta T_j| $. Given a frontier $ \delta T_j $, there can be no more labelings than there are Motzkin paths, of which there are $ O(3^{|\delta T_j|}) $. Thus if we let $ \kappa = \max_j\{ |\delta T_j| \} $, there are $ O(\ell k 3^\kappa) $ possible triples. Each triple has to loop through no more than $ \beta^2 $ successive counts, for a complexity of $ O(\beta^2 \ell k 3^\kappa) $.
	
	The space complexity of this algorithm is also high. It is the number of possible triples,
	$ O(\ell k 3^\kappa) $, which is not something that can always fit into memory. For this reason some external database or distributed storage may be required for larger inputs, and the IO operations may slow the algorithm down further in practice.

    If we take $\kappa$ to be the exact cutwidth of $G_0$, both time and space complexity are $\ell k 3^{O(\sqrt{n})}$ for planar graphs of bounded degree. Computing the cutwidth exactly can, for this case, be done using the $O(n^{\kappa-1})$ algorithm of \cite{cutwidthFPT} for an algorithm running in time $\ell k 2^{O(\sqrt{n})} + 2^{O(\sqrt{n}\log n)}$.
	\subsubsection{Extensions for Algorithmic Redistricting}\label{sec:ext}
	
	In this section, we describe how to add dimensions to the algorithm to satisfy some criteria required in the application of algorithmic redistricting. The algorithm above uniformly samples from \emph{all} possible connected $ k $-partitions of $ G $, of which the vast majority do not form reasonable districting plans because they do not meet the compactness and population constraints typically required in political redistricting. To this end, we can modify the algorithm to count the number of feasible $ k $-partitions.
	
	\textbf{Compactness:} First, we look at the compactness criteria. This can come in many forms, such as the Polsby-Popper score, the Reock test, some measures of discrete compactness, and many more. Each have quite a few issues and inconsistencies, but we will use the discrete perimeter, which is defined to be the size of the cut partitioning the graph into $ k $ parts, or the length of the SAW. This compactness measure is simple to incorporate into the algorithm by just adding an additional parameter $ i $ for the length of the partial SAW. The update rules then increase $ i $ by $ 2 $ whenever a $ 2 $-$ 3 $ pair is introduced or a split occurs, and by $ 1 $ when a labeled edge continues through the given face $ f_j $, or an entry occurs. For each case, it is not difficult to see by how much the length of the partial SAW increases. 
	
	We will refer to the compactness bound as $ c $. Both space and time complexity increase with this extension by $c$, for a total complexity of $O(c\ell k 3^\kappa)$. 
	
	\textbf{Population:} More difficult to incorporate is the population balance criterion. We sketch a method to keep track of the population balance during enumeration, although the added complexity was too great for implementation, so we used rejection sampling instead. For each district, the population balance can be measured by the deviation from the mean population of a district. The legal requirements are prohibitively tight, as mapmakers usually cut up census blocks to reach these requirements. As typically done in the algorithmic redistricting literature \cite{ReCom}, 
	we significantly loosen the criterion of population balance, and only ask that the maximum deviation of any district from the mean is bounded by some fixed percentage.
	
	Then if the total population of a state is $ P $, we introduce $ k $ new dimensions to the DP table, one  for each district, each with $ P $ possible values. There is then also a requirement to keep track of which part of the frontier belongs to which district, so we can know how to update the population counts. To do so for a given labeling along a given frontier, consider the tree formed by creating a node for each section of the frontier that corresponds to a possible district created by this SAW. A node is created for sections separated by $1$'s, and sections enclosed by $2$-$3$ pairs. The edges of the tree are then between neighboring sections, and sections that directly enclose each other. An assignment of these sections to the $k$ districts is exactly a $k$-coloring of this tree. Thus it suffices to keep track, for each labeling of each frontier, of a $k$-coloring of the above tree, and the populations of each district. The number of $k$-colorings of a tree on $n$ vertices is $k(k-1)^n=O(k^n)$, and the number of nodes of this tree is bounded by $\kappa$, the max length of a frontier. Thus the added complexity of this method to keep track of population is $O(P^{k}k^{\kappa})$, which is prohibitively large for most practical cases.
 \dnote{I could talk about why binning population counts doesn't really work, but it's just a sketch} \jnote{I'm not sure what the binning does but it sounds relevant to address that you considered it. Maybe write it up, and we can comment it out later if necessary.}
	
	A natural idea to improve this complexity would be to bin population counts into bins of size $p$. 
 There are some difficulties with this approach. First, 
    the error factor would grow with the number of faces $\ell$, so population counts could only be guaranteed up to a factor of $\ell p$. Second, $p$ must be no greater than the minimum population of any precinct, already implying $P/p\geq n$. Binning would then only be effective when the number of districts $k$ is constant, or at most $O(\sqrt{n})$. This approach gives a tradeoff between complexity and accuracy. For example, if we wanted to bound population deviation by some percentage $0\leq x\leq 1$ from the mean district, we might take some error such that $\ell p = \frac{1}{2} x P/k$, which means $P/p= 2k\ell/x$. Then, to get a uniform distribution, sample from partitions with binned population deviation bounded by $\frac{3}{2} x P/k$, and using rejection sampling accept only those where the true population is within $xP/k$. Intuitively, at least half of the samples should be good. The added complexity would then be $O((2k\ell/x)^k k^\kappa ),$ which is $O(n^k k^{\kappa})$, where $\kappa=O(\sqrt{n})$ when $k$ is constant for bounded degree planar graphs with bounded faces.
    	

 	\textbf{Sampling valid districts:} Putting all these together, our algorithm to uniformly sample valid districting plans proceeds by computing the exact cutwidth of $ G_0 $ in $ O(n^{\kappa-1}) $ time, and then runs the dynamic programming routine with the compactness and population extensions in $ O(c \ell k 3^{\kappa} (P/p)^k k^{\kappa}) $, with an error factor depending on $p$ and $\ell$. As the cutwidth of a bounded degree planar graph is $ O(\sqrt{n}) $, and $ \ell $ is within a constant factor of $ n $ in this case, the cumulative time complexity is 
  \dnote{Big O notation is a little awkward throughout, as big O in exponents doesn't jump down. How should I write these running times nicely?}
        \[ 2^{O(\sqrt{n}\log n)} + 2^{O(\sqrt{n}\log k)} c n (P/p)^k. \]
 
	\section{Creating a Traversal Order}\label{sec4}
	
	The exponential factor $ \kappa $ in the complexity is controlled by the ordering of the faces $ \{T_j\}_{j=1}^\ell $.
	In this section, we give a heuristic algorithm to compute $\kappa$, that is the cutwidth of the primal graph, via its dual. The algorithm intends to perform well on grid-like graphs, making it suitable for the practical application of redistricting. \dnote{This whole section is pretty pointless with all the known methods of computing cutwidth}

	
	We seek to find an ordering of the faces of $G$ $ \{T_j\}_{j=1}^\ell $ that minimizes $ \max_j\{|\delta T_j|\} $, and to this end present a recursive algorithm that performs well in practice. However, the algorithm has no theoretical backing and performs poorly on certain graphs.
	
	The algorithm sets up some intermediate frontiers for each vertex in $ \tau $ using a simple dynamic programming algorithm that minimizes the maximum length of a sequence of intermediate frontiers, and then recurses on the slices between each frontier. It additionally keeps track of traversed edges, and indexes the outer face of each recursive call such that the initial edge is safe. Without loss of generality assume also that each vertex of $ \tau_G $ has degree at least $ 3 $ for simplicity. If this is not the case, we can replace degree-two vertices by an edge. It is described in full in Algorithm \ref{alg:trav}.
	
	\begin{algorithm}[h]
		\caption{The algorithm to generate a traversal order given the dual graph $ G $ and a starting edge $ e $ defined by the indexing of $ \tau_G $ and a path over traversed ``safe" edges $ p_s = (V_s, \tau_s) $.}\label{alg:trav}
		\begin{algorithmic}
			\Function{Generate\_Traversal}{$ G = (V,E) $, $ p_s = (V_s, \tau_s) $}
			\If{$ G $ only has one nontrival non-outer face $ f $}
			\Return $ f $.
			\Else{ Let $ \tau_G \supseteq \tau_s $ be the outer face of $ G $.}
			\EndIf
			\If{$ |\tau_G| = 3 $}
			\State Add an arbitrary face connected to a safe edge to the traversal.
			\EndIf
			\State Let $ V_s' $ be $ V_s $ with the path ends removed and $ V_{\tau_G} $ the vertices of $ \tau_G $.
			\State $ G'_{u,v} \gets (V \setminus (V_{\tau_G} \setminus (V_s' \cup \{u,v\})), E\setminus \tau_G) $.
			\State Compute all shortest paths $ P(u,v) $ for $ u,v\in V $ in $ G' $ using Dijkstra's algorithm.
			\State $ u_{start}\gets (\tau_G)_0, $ $ v_{start}\gets (\tau_G)_{|\tau_G|-1} $.
			\State $ d(u_{start}, v_{start}) = |P(u_{start},v_{start})| $.
			\For{$ \ell=3 \ldots |\tau_G|-1 $}
			\For{$ i=0 \ldots \ell-1 $}
			\State $ j\gets |\tau_G| - \ell + i $.
			\State $ u\gets (\tau_G)_i, $ $ v\gets (\tau_G)_j $.
			\State $ u'\gets (\tau_G)_{i+1}, $ $ v'\gets (\tau_G)_{j-1} $.
			\State $ v_{pen} \gets \begin{cases}
				0 \quad \text{if }(v',v)\in \tau_s\\
				1 \quad \text{otherwise}
			\end{cases} $.
			\State $ u_{pen} \gets \begin{cases}
				0 \quad \text{if }(u,u')\in \tau_s\\
				1 \quad \text{otherwise}
			\end{cases} $.
			\If{$ i=0 $}
			\State $ d(u, v) \gets d(u, v') + v_{pen} $.
			\State $ t(u, v) \gets 0 $.
			\ElsIf{$ j=|\tau_G|-1 $}
			\State $ d(u, v) \gets d(u', v) + u_{pen} $.
			\State $ t(u, v) \gets 1 $.
			\Else
			\State $ d(u,v) \gets \min\big\{ \max\{ d(u,v') + v_{pen}, |P(u,v')| \}, \max\{ d(u',v) + u_{pen}, |P(u',v)| \} \big\} $.
			\State $ t(u, v) \gets \argmin \big\{ \max\{ d(u,v') + v_{pen}, |P(u,v')| \}, \max\{ d(u',v) + u_{pen}, |P(u',v)| \} \big\} $.
			\EndIf
			\EndFor
			\EndFor
			\State $ u_{end}, v_{end} \gets \argmin\{ d(u,v) : (u,v)\in \tau_G, (u,v) \neq (u_{start},v_{start}) \} $.
			\State $ u,v \gets u_{end}, v_{end} $.
			\For{$ j=1 $ to $  $}
			\State Let $ u',v' $ be the step from $ u,v $ defined by the traceback function $ t $.
			\State $ H \gets $ the subgraph of $ G $ enclosed by $ P(u, v) $ and $ P(u', v') $, including the paths.
			\State Add \Call{Generate\_Traversal}{$ H, P(u,v) \cup (\tau_s \cap P(u', v')) $} to the traversal.
			\State $ u,v \gets u',v' $.
			\EndFor
			
			\Return The traversal.
			\EndFunction
		\end{algorithmic}
	\end{algorithm}
	
	
	
	
	Algorithm \ref{alg:trav} terminates after at most $|F|$ recursive calls, as each pair of intermediate traversals $P(u,v), P(u',v')$ have some face between them, and because the outer face satisfies $|\tau_G|>3$, no subgraph $H$ will be the full graph. 
	Also $ P(u,v) $ and $ P(u',v) $ meet only once and then are the same, because Dijkstra's 
 algorithm stores the same shortest path for both at the vertex where they meet. Thus the subgraph $H$ is connected as well.
	
	One important feature of this algorithm is that it also returns a traversal of the form we defined, where the frontier is contiguous. General algorithms for cutwidth do not ensure that the ordering is such that the cut remains connected in the dual as we add the vertices one by one. 
	
	
	\section{Experimental Results}
	
	Our theoretical results are backed by an implementation of our sampling algorithm. We ran two experiments: one to test the computational limits of the sampling algorithm using a two-district scenario, and another to demonstrate its effectiveness at generating realistic partitions in a three-district scenario, which is more realistic for redistricting.
	
	The experiments were run on an 18-core 3GHz Intel Xeon W-2295 CPU, with 256 GB of RAM.
	
\begin{figure}[h]
		\centering
		\begin{tabular}{ccccc}
			\subfloat{\includegraphics[clip,trim={1.7in} {.6in} {1.4in} {.65in}, scale=.25]{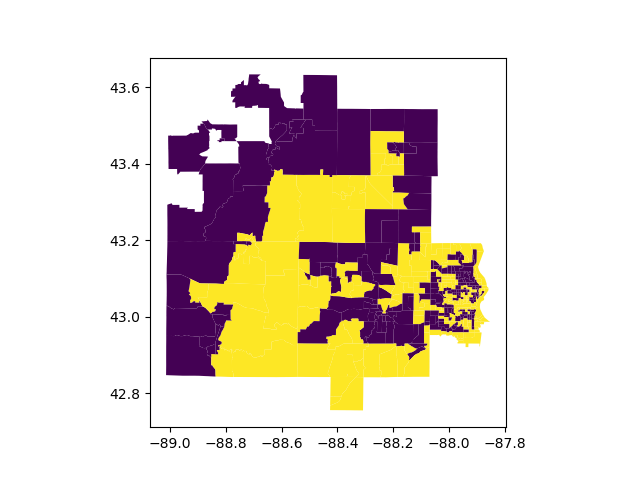}} &
			\subfloat{\includegraphics[clip,trim={1.7in} {.6in} {1.4in} {.65in}, scale=.25]{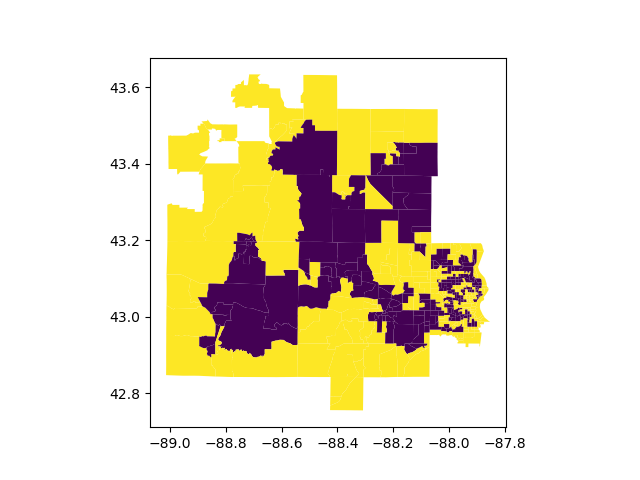}} &
			\subfloat{\includegraphics[clip,trim={1.7in} {.6in} {1.4in} {.65in}, scale=.25]{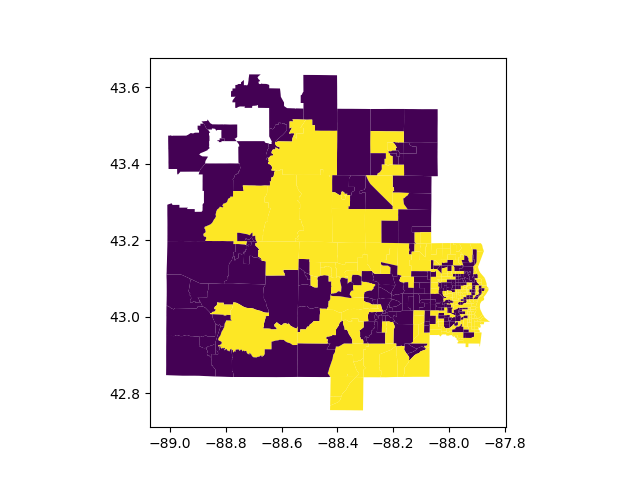}} &
			\subfloat{\includegraphics[clip,trim={1.7in} {.6in} {1.4in} {.65in}, scale=.25]{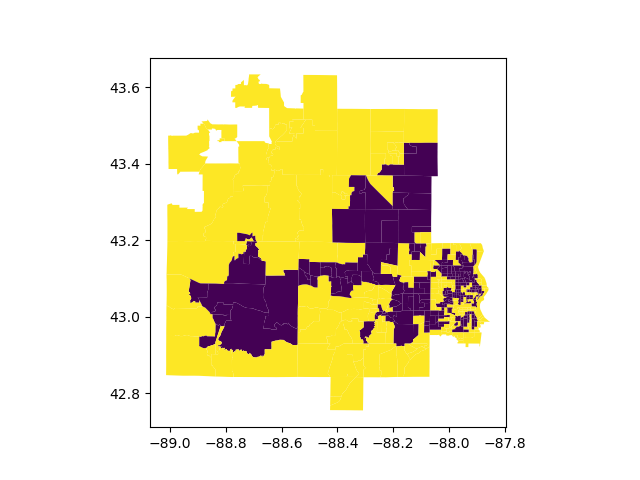}} &
			\subfloat{\includegraphics[clip,trim={1.7in} {.6in} {1.4in} {.65in}, scale=.25]{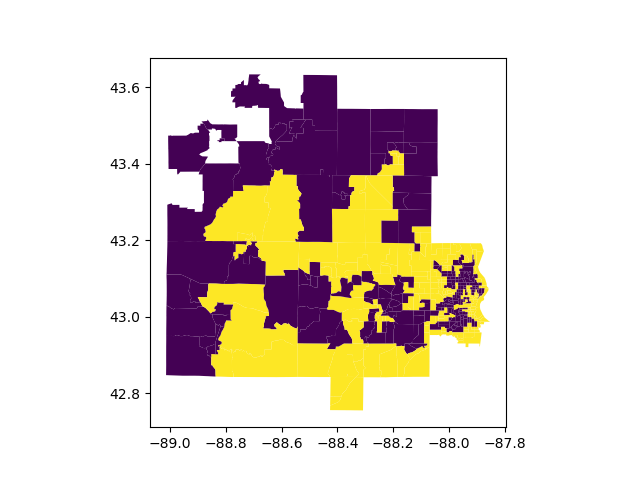}} \\
			\subfloat{\includegraphics[clip,trim={1.7in} {.6in} {1.4in} {.65in}, scale=.25]{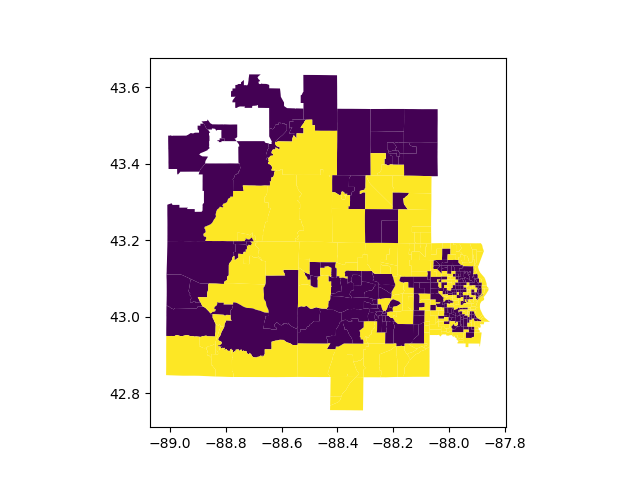}} &
			\subfloat{\includegraphics[clip,trim={1.7in} {.6in} {1.4in} {.65in}, scale=.25]{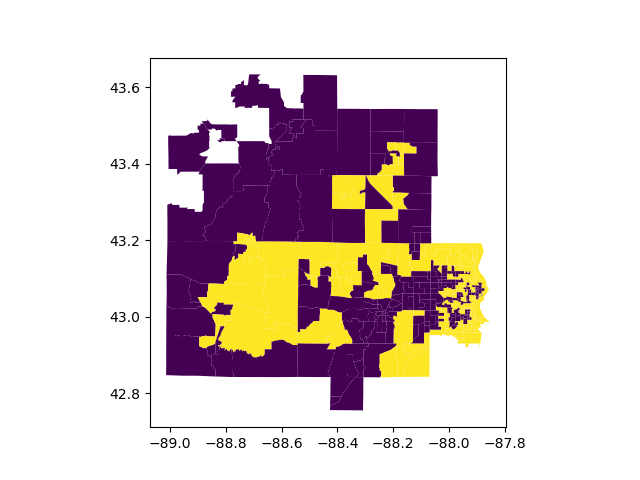}} &
			\subfloat{\includegraphics[clip,trim={1.7in} {.6in} {1.4in} {.65in}, scale=.25]{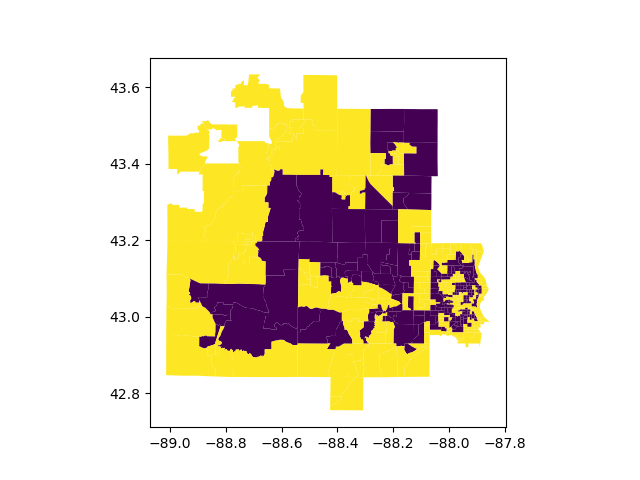}} &
			\subfloat{\includegraphics[clip,trim={1.7in} {.6in} {1.4in} {.65in}, scale=.25]{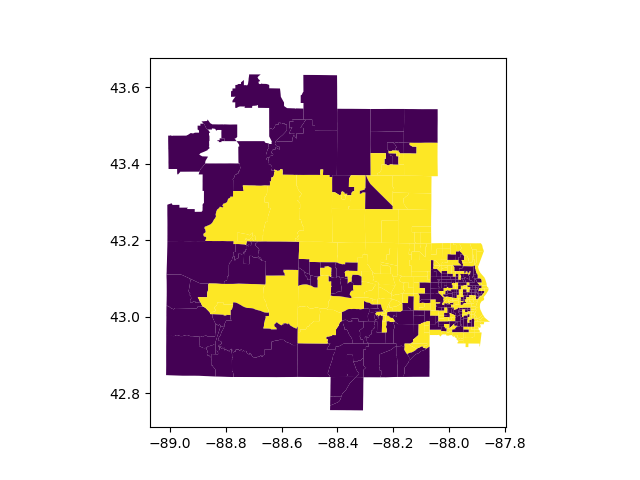}} &
			\subfloat{\includegraphics[clip,trim={1.7in} {.6in} {1.4in} {.65in}, scale=.25]{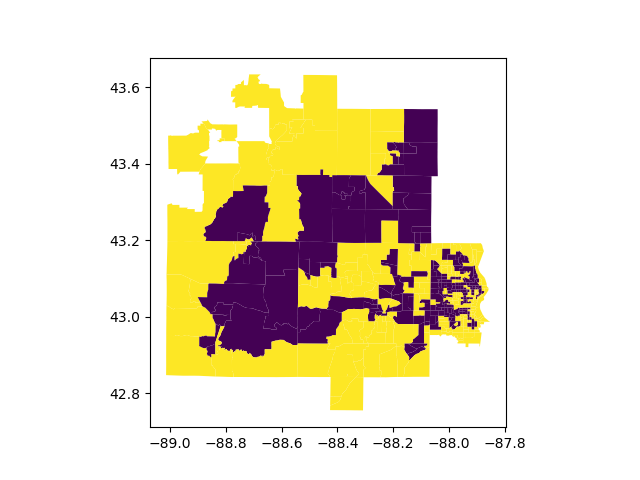}}
		\end{tabular}
		\caption{Randomly generated bipartitions of the census tracts of the union of Wisconsin congressional districts 4 and 5 without compactness constraints.}\label{fig:maps1}
	\end{figure}

	For the first experiment, we generated bipartitions of the union of congressional districts 4 and 5 in Wisconsin. These districts were taken from the enacted 2022 redistricting plan \cite{Wisconsin2022}. 
	This union was composed of $ 1263 $
	census block groups (CBGs) and $ 421 $ census tracts. 
	Districts are usually built from CBGs, but we were only able to use the census tract level, where Algorithm \ref{alg:trav} generated a traversal with $ \kappa=42 $. 
	For comparison, we first generated samples without any compactness bounds, with population correctness enforced by rejection sampling. Some of these samples are shown in Figure \ref{fig:maps1}. The count of \emph{all} possible such bipartitions was approximately $ 5.532338418\times 10^{80} $. 
	
\begin{figure}[h]
		\centering
		\begin{tabular}{ccccc}
			\subfloat{\includegraphics[clip,trim={1.7in} {.6in} {1.4in} {.65in}, scale=.25]{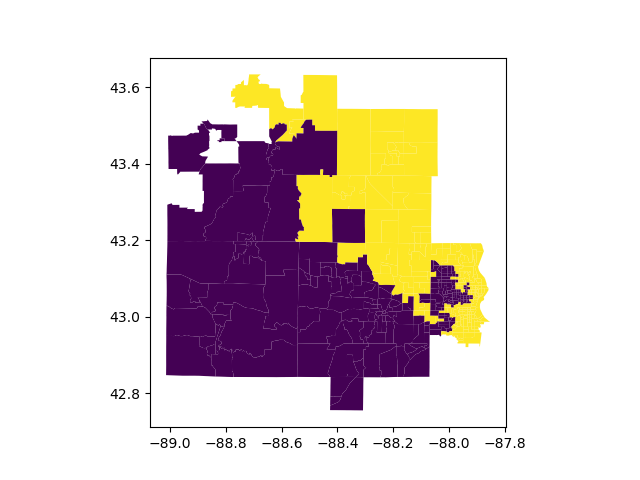}} &
			\subfloat{\includegraphics[clip,trim={1.7in} {.6in} {1.4in} {.65in}, scale=.25]{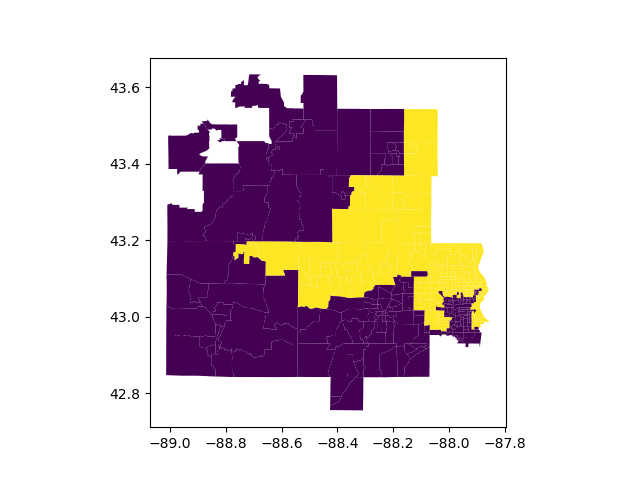}} &
			\subfloat{\includegraphics[clip,trim={1.7in} {.6in} {1.4in} {.65in}, scale=.25]{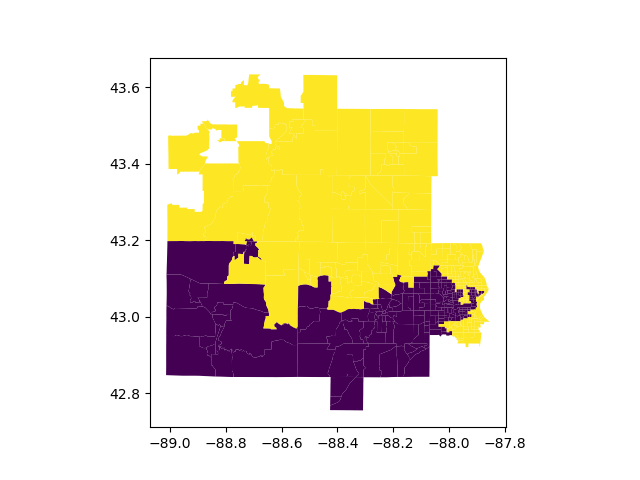}} &
			\subfloat{\includegraphics[clip,trim={1.7in} {.6in} {1.4in} {.65in}, scale=.25]{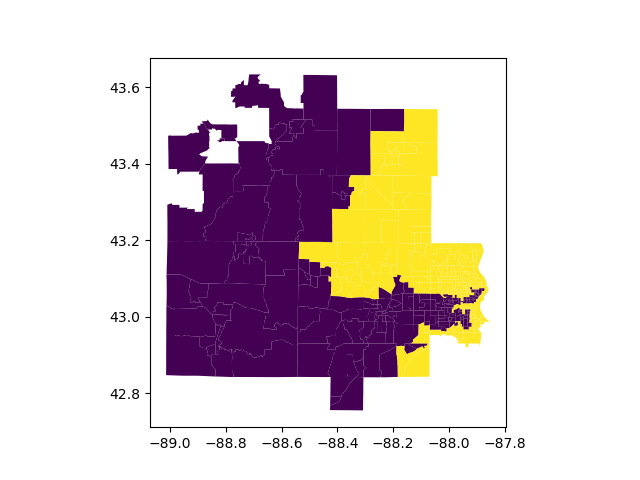}} &
			\subfloat{\includegraphics[clip,trim={1.7in} {.6in} {1.4in} {.65in}, scale=.25]{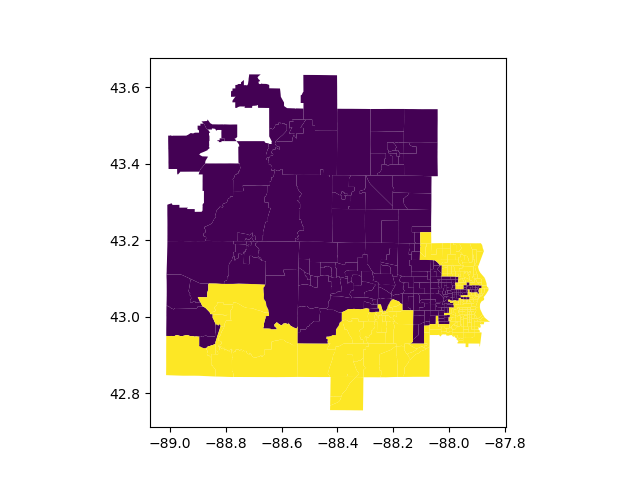}} \\
			\subfloat{\includegraphics[clip,trim={1.7in} {.6in} {1.4in} {.65in}, scale=.25]{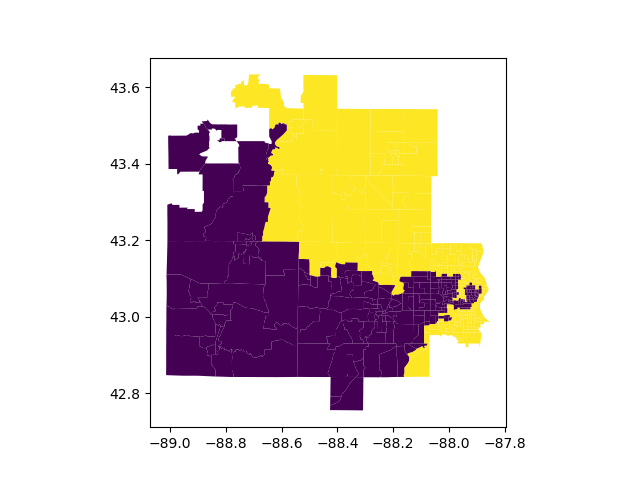}} &
			\subfloat{\includegraphics[clip,trim={1.7in} {.6in} {1.4in} {.65in}, scale=.25]{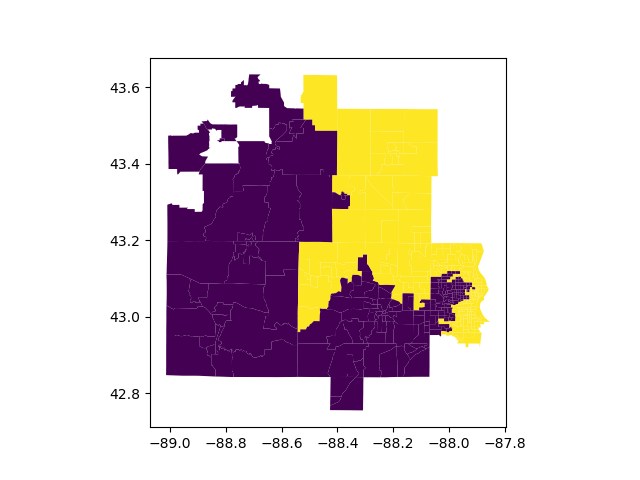}} &
			\subfloat{\includegraphics[clip,trim={1.7in} {.6in} {1.4in} {.65in}, scale=.25]{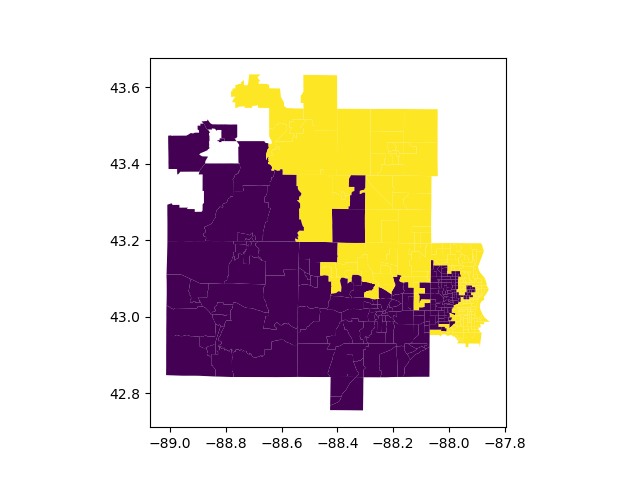}} &
			\subfloat{\includegraphics[clip,trim={1.7in} {.6in} {1.4in} {.65in}, scale=.25]{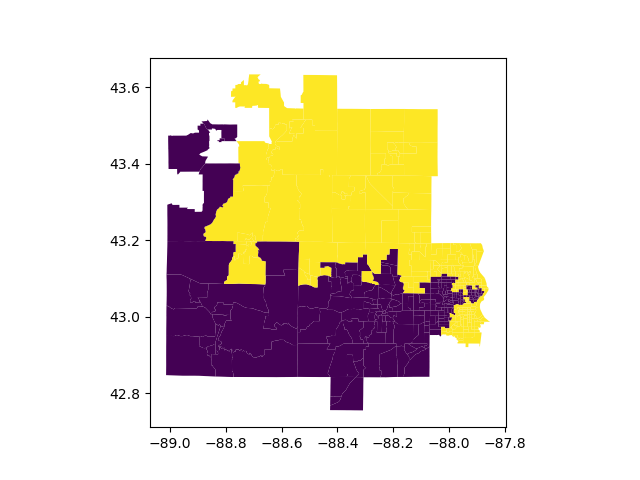}} &
			\subfloat{\includegraphics[clip,trim={1.7in} {.6in} {1.4in} {.65in}, scale=.25]{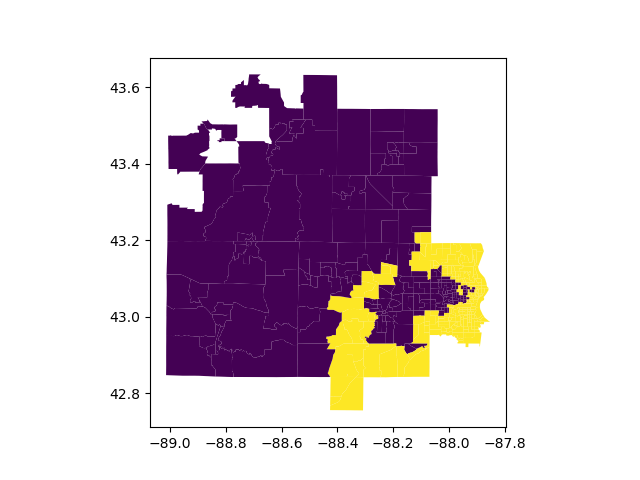}}
		\end{tabular}
		\caption{Randomly generated bipartitions of the census tracts of the union of Wisconsin congressional districts 4 and 5 with added compactness constraints.}\label{fig:maps2}
	\end{figure}

	We then sampled from this space with compactness constraints. To accomplish this, we enforced a discrete perimeter constraint on the number of edges between the two sampled districts, limiting the length of the non-self-intersecting path to $ 150 $ or less. This constraint, as described in Section \ref{sec:ext}, could be integrated into the counting algorithm, and only the population constraints were enforced via rejection sampling. 
	However, as this change increased the time and space complexity by a factor of $ 150 $, additional speedups were necessary. This was done by an additional bound of $ 4 $ on 
	the maximum width of a Motzkin path on the frontier, i.e., a bound on the number of times the non-self-intersecting path can intersect any given frontier. An intersection here is a $ 2 $-$ 3 $ pair, rather than a $ 1 $ in the notation of Section \ref{sec:alg}. This is a reasonable constraint for compact districts, but is unfortunately traversal dependent, which means it is not a good way to define the space of valid districting plans we are sampling from.
	We then counted approximately $ 2.809736245\times 10^{35} $ such bipartitions, without the population constraints. Samples with population constraints are shown in Figure \ref{fig:maps2}.
	
\begin{figure}[h]
		\centering
		\begin{tabular}{cccc}
			\subfloat{\includegraphics[clip,trim={1in} {1.1in} {1in} {1.2in}, scale=.25]{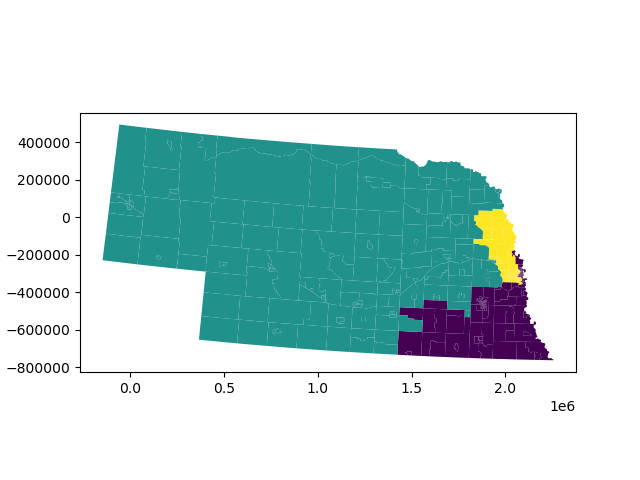}} &
			\subfloat{\includegraphics[clip,trim={1in} {1.1in} {1in} {1.2in}, scale=.25]{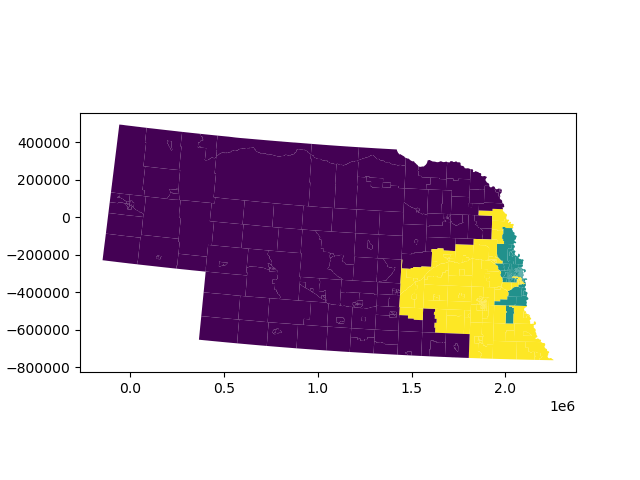}} &
			\subfloat{\includegraphics[clip,trim={1in} {1.1in} {1in} {1.2in}, scale=.25]{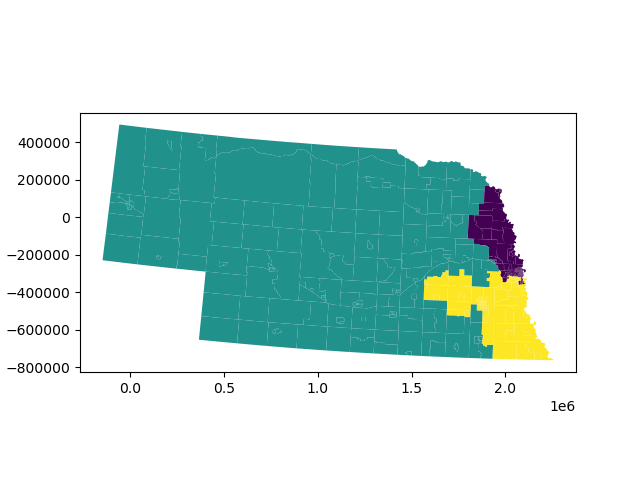}} &
			\subfloat{\includegraphics[clip,trim={1in} {1.1in} {1in} {1.2in}, scale=.25]{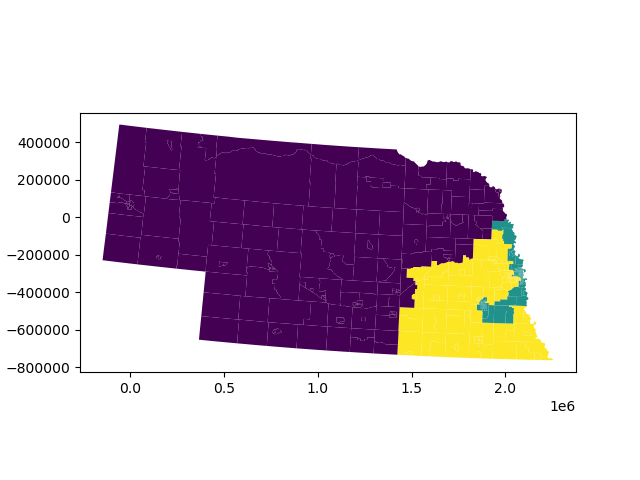}} \\
			\subfloat{\includegraphics[clip,trim={1in} {1.1in} {1in} {1.2in}, scale=.25]{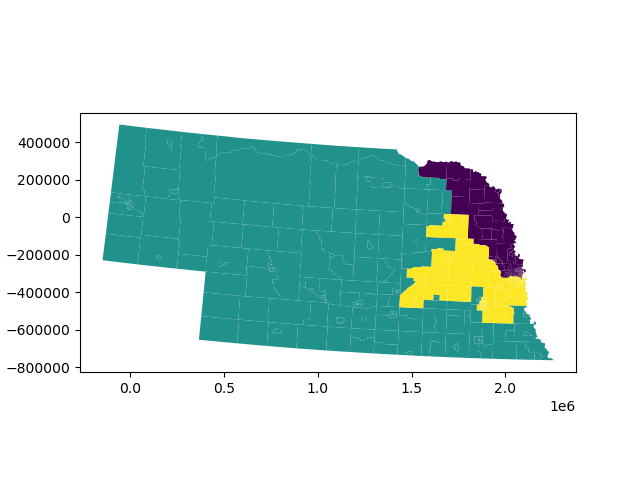}} &
			\subfloat{\includegraphics[clip,trim={1in} {1.1in} {1in} {1.2in}, scale=.25]{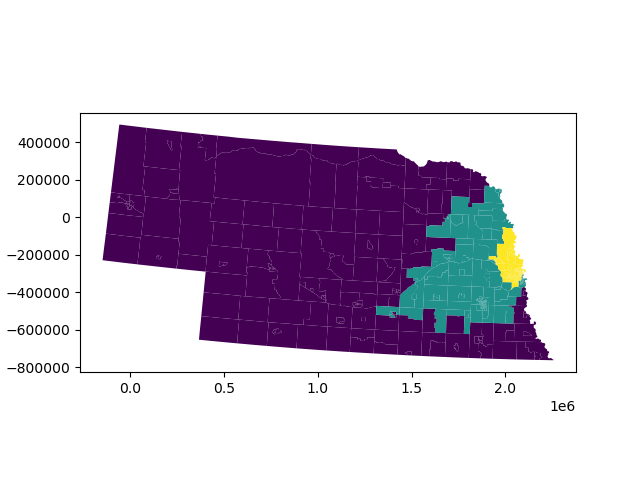}} &
			\subfloat{\includegraphics[clip,trim={1in} {1.1in} {1in} {1.2in}, scale=.25]{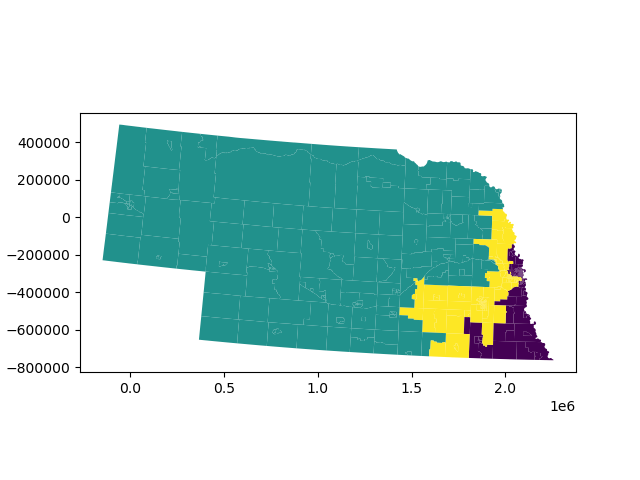}} &
			\subfloat{\includegraphics[clip,trim={1in} {1.1in} {1in} {1.2in}, scale=.25]{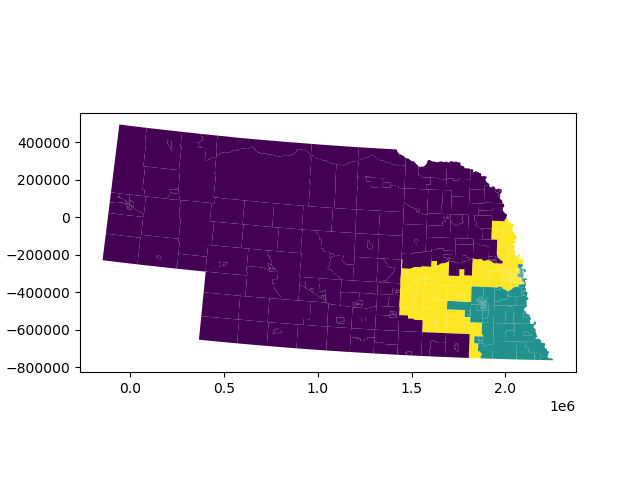}} \\
			\subfloat{\includegraphics[clip,trim={1in} {1.1in} {1in} {1.2in}, scale=.25]{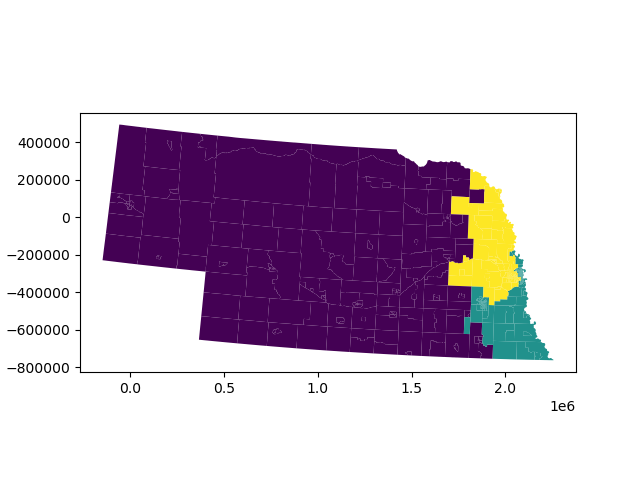}} &
			\subfloat{\includegraphics[clip,trim={1in} {1.1in} {1in} {1.2in}, scale=.25]{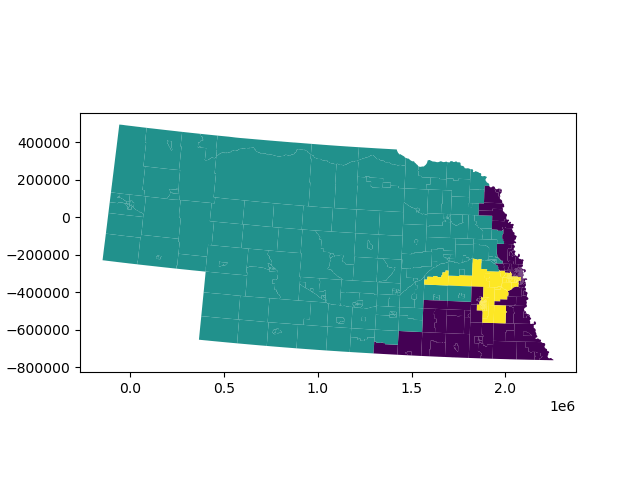}} &
			\subfloat{\includegraphics[clip,trim={1in} {1.1in} {1in} {1.2in}, scale=.25]{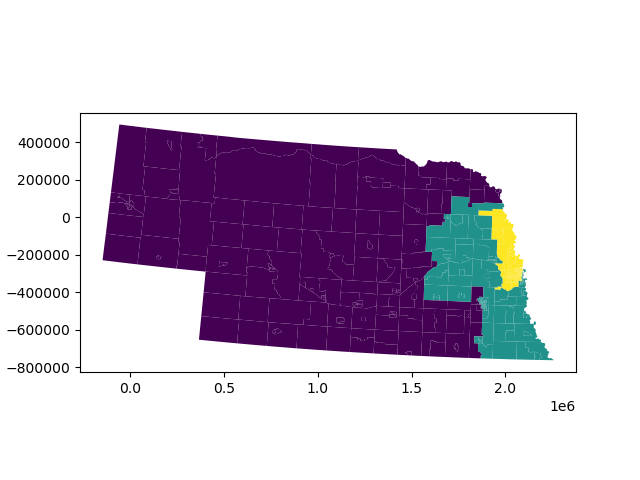}} &
			\subfloat{\includegraphics[clip,trim={1in} {1.1in} {1in} {1.2in}, scale=.25]{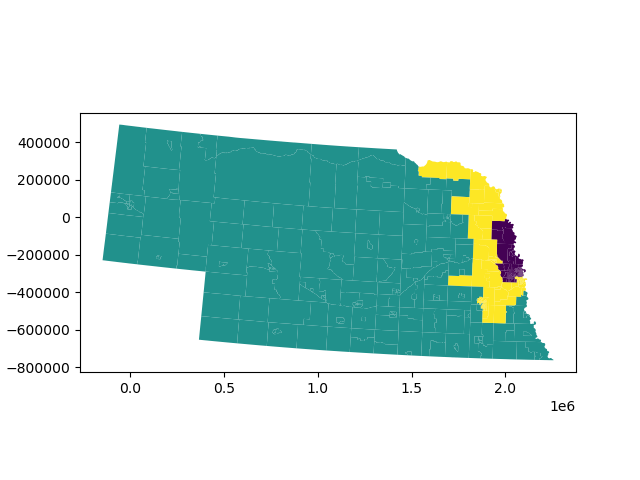}}
		\end{tabular}
		\caption{Randomly generated three partitions of the census tracts of Nebraska.}\label{fig:maps3}
	\end{figure}

	The second experiment aims to construct a districting plan for an entire state.  For this goal, a 3-partition of the census tracts of Nebraska proved to be the suitable choice, as it has 3 congressional districts and admitted a shorter traversal. 
	For this dataset, $ n=553 $, and algorithm \ref{alg:trav} found a traversal with $ \kappa=39 $. We imposed a discrete perimeter constraint of $ 100 $, and a rather restrictive bound of $ 2 $ to the maximum width of a Motzkin path (i.e., the number of times the path could ``loop" across any given frontier was limited to 2). With these constraints, we counted around $ 3.788915905\times 10^{18} $ partitions. Upon sampling, we imposed a 
	population constraint of $ 15\% $, meaning that the population of each district must be within $ 15\% $ of the ideal population, namely the total population of the state divided by $ 3 $. Out of $ 50,000 $ samples, $ 19 $ satisfied the population constraints as well, $ 12 $ of which are shown in Figure \ref{fig:maps3}.
	
	\subsection{Discussion}
	These counts confirm that the number of valid districting plans is incredibly large. They also show that when considering all partitions of a graph, an exponential proportion of them do not satisfy population and compactness constraints. In our example, only about one out of $10^{45}$ partitions satisfied \emph{just the compactness constraints}. \dnote{Could put charts showing how counts grow with compactness to back this up}\jnote{yes!}
	
	The results also demonstrate the computational limits of this method. The computed $ \kappa $ values where greater than the $ O(\sqrt{n}) $ bound. This could be due to the weaknesses of our cutwidth algorithm, choice of data set, or the impact of constant factors in the theoretical bound. Each experiment was scaled to be near the limit of our machine, with runs taking up to 24 hours \dnote{check this}, and space usage of up to $ 2.4 $ terabytes.
	
	\dnote{Should there be a section for comparison to other work? How would we compare this to existing algorithms, besides what's discussed in the introduction?}

	\section{Future Work}\label{future}
	
	In this paper we gave an algorithm 
 to uniformly sample the space of \emph{all} possible $ k $-partitions of a graph. We extended this to sample from the space of all \emph{compact} partitions, where compactness is measured using the discrete perimeter. It remains to further restrict this space to all \emph{valid} districting plans, which would enforce population constraints as well. Another step would be to uniformly sample \emph{realistic} districting plans, where constraints such as splitting rules or the Voting Rights Act would be enforced as well. This would require an appropriate mathematical definition of the sample space, which is not necessarily possible.
	
	
	As for the computational complexity, our experimental results have shown that the subexponential running time can be feasible for some granularity of the maps. Cleaning the TM algorithm would allow for a more efficient implementation, and increase the practical limit for $ \kappa $ by 6 to 10. Using more efficient methods to compute cutwidth and allowing a noncontiguous boundary would also decrease $ \kappa $ and allow for sampling from larger datasets.
	Nonetheless we are 
	able to enumerate all the $ k $-partitions of a graph, and with this generate the first sample that truly includes \emph{every} valid districting plan.

	\bibliographystyle{plain}
	\bibliography{gerrypaper.bib}
	
\end{document}